# Observations of the Cosmic Microwave Background

C. L. Bennett[a] *

[a]e-mail: bennett@stars.gsfc.nasa.gov
Code 685, Infrared Astrophysics Branch, Laboratory for Astronomy and Solar Physics
NASA/GSFC, Greenbelt, MD 20771, U.S.A.

The properties of the cosmic microwave background radiation provide unique constraints on cosmological models, i.e. on the content, history, and evolution of the Universe. I discuss the latest measurements of the spectral and spatial properties of the cosmic microwave background radiation. Recent measurements from NASA's *Cosmic Background Explorer* (*COBE*) satellite, and from balloon-borne and ground-based platforms, are summarized and their cosmological implications are discussed.

## 1. INTRODUCTION

It is remarkable that the few cosmological observables available to us place severe constraints on acceptable cosmological models. Probably the most fundamental of the cosmological observables is the recession of the galaxies, i.e. the Hubble expansion of the universe [1,2]. The discovery that our universe is expanding led directly to the notion of a past epoch when the universe was hot and dense; the big bang theory follows as a natural consequence of the expansion. The ages of the oldest stars clusters in the halo of the Milky Way galaxy are estimated to be $16 \pm 2$ billion years old [3], roughly in agreement with the age of the universe deduced from the Hubble expansion. Within the context of the big bang theory, calculations of nucleosynthesis make predictions for the abundances of the light chemical elements that agree well with the observed abundances, and the fact that there are exactly three flavors of neutrinos [4–6]. A cosmic microwave background (CMB) radiation was predicted based on the big bang theory, and was discovered in 1964 [7,8] and its predicted blackbody spectral shape was later confirmed in detail [9].

The hot big bang theory has passed stringent tests, but it is only a cosmological framework. It says little to nothing about the initial conditions, or levels of perturbations in the universe, or about the material content of the universe. The growth of the particular patterns and velocities of material structures must also be explained. Thus further observables are required to shape models of galaxy formation and evolution. The spatial distribution and velocities of the galaxies probe the gravitational field on large scales. Data on the time evolution of galaxy luminosities, spectra, clustering, and densities help in our understanding of their formation and growth.

In part because of the high degree of CMB isotropy, a new cosmological paradigm arose where a large fraction of the mass in the universe is nonbaryonic dark matter that does not couple in any way to light. The nonbaryonic dark matter could begin gravitational potential growth while the baryonic matter remained constrained by the pressure from the CMB radiation field. In this way the observed luminous structures in the sky could be made consistent with much smaller CMB temperature fluctuations.

A great advance in the characterization of the CMB properties came about recently due to NASA's *COBE* mission. The satellite and its instruments were designed, built, managed, and

*I acknowledge my colleagues on the *COBE* Science Working Group and all of those who support the scientific endeavor, especially the Astrophysics Division of NASA Headquarters. NASA/GSFC is responsible for the design, development, and operations of the *COBE*. Scientific guidance is provided by the *COBE* Science Working Group. GSFC is also responsible for the development of the analysis software and the delivery of the mission data sets.

launched by the NASA Goddard Space Flight Center. *COBE* and its instruments worked beautifully. The Far Infrared Absolute Spectrophotometer (FIRAS) instrument made a precision measurement of the spectrum of the CMB from 1 cm to 100 $\mu$m, providing an excellent match between the CMB spectrum and that predicted by the simple big bang model. The Differential Microwave Radiometers (DMR) instrument searched for CMB anisotropies on angular scales larger than 7° at frequencies of 31.5, 53, and 90 GHz, providing the first detection of temperature fluctuations. In this paper I discuss the observational results and cosmological implications from *COBE* and other recent CMB experiments, which probe anisotropies over a wide range of wavelengths and angular scales.

## 2. THE CMB SPECTRUM

### 2.1. Introduction

The detection of interstellar CN molecular absorption lines by Adams [10] and the subsequent conclusion by McKellar [11] that these lines were consistent with the CN being in a radiation bath of temperature 2.3 K was, in retrospect, the first detection of the CMB radiation and determination of its temperature. A quarter of a century later Penzias and Wilson [7] discovered the radiation, also by accident, via its microwave continuum. Their temperature measurement of 3.5 ± 1.0 K at a wavelength of 7 cm was interpreted by Dicke et al. [8] as being the afterglow of the big bang. An upper limit of about 10% was set on the anisotropy of the microwave emission.

Thirty years after the Penzias and Wilson measurement, the CMB is now generally accepted to be the afterglow radiation from a hot and dense epoch in the early universe. Since the annihilation of positrons at $z \approx 3 \times 10^8$, the CMB photons have outnumbered protons, neutron, and electrons by a factor of $\sim 10^9$. The number of CMB photons is fixed by $z \approx 3 \times 10^6$, about a year after the big bang, when the double quantum and free-free processes slow relative to the cosmic expansion [12]. After that time the CMB spectrum could deviate from a blackbody form only if energy is added [12–15]. At first, multi-ple Compton scattering is sufficient to establish a pseudo-equilibrium form, i.e. a Bose-Einstein spectrum with the number density of photons at energy $\epsilon$ given by $N(\epsilon) = 1/(e^{\epsilon/kT+\mu} - 1)$, where $\mu$ is the dimensionless chemical potential, and $k$ is the Boltzmann constant. After $z \approx 10^5$, this process also becomes slow relative to the cosmic time scale, and the CMB spectrum can be distorted in other ways. One likely form is called a Compton distortion, which is equivalent to mixing blackbodies of differing temperatures. The Compton distortion is usually parameterized by $y = (\sigma_T/m_e c^2) \int n_e k(T_e - T_{CMB}) cdt$, where $\sigma_T$ is the Thomson scattering cross-section, $m_e$ is the electron mass, $n_e$ is the electron density, $T_e$ is the electron temperature, $T_{CMB}$ is the CMB temperature, $c$ is the speed of light, and $cdt$ is a distance element along the line of sight; thus $y$ is proportional to the electron pressure integrated along the line of sight. A Compton distortion of the spectrum can become important when $(1+z)d\tilde{y}/dz > 1$, where $\tilde{y} \equiv (\sigma_T/m_e c^2) \int n_e k T_e cdt$, which occurs $\sim$ 2000 years after the big bang. The thermodynamic temperature distortion observed at a frequency $\nu$ is

$$\frac{\delta T}{T} \approx y \left( x \frac{e^x + 1}{e^x - 1} - 4 \right), \quad (1)$$

where $x = h\nu/kT_{CMB}$, and $h$ is the Planck constant [15].

### 2.2. FIRAS Results

The most recent FIRAS results show that the CMB spectrum deviates from that of a blackbody by less than 0.03% of the peak intensity over the wavelength range from 0.5 to 5 mm [9], implying that the radiation was once optically thick and in equilibrium with matter at a single temperature, and that no physical processes in the history of the universe since that epoch have been energetic enough to perturb the spectrum. Figure 1, from [9], shows these results. The residuals are shown after subtraction of the best fit combination of a blackbody, dipole, and galactic emission. The residuals are for the case where $y = \mu = 0$. The weighted rms residual is only 0.01% of the peak brightness.

The $y$ and $\mu$ curves in Figure 1 show the shapes of distortion that would be produced if these spectral distortion parameters have the 95% confidence limit values of $|y| = 2.5 \times 10^{-5}$ and $|\mu| = 3.3 \times 10^{-4}$. If the CMB spectrum is a graybody, then its emissivity is limited to $1 \pm 0.00041$ (95% CL). These spectral results are a thousand times better than was known before *COBE*. The FIRAS also showed that the cosmic dipole has the expected spectrum [16].

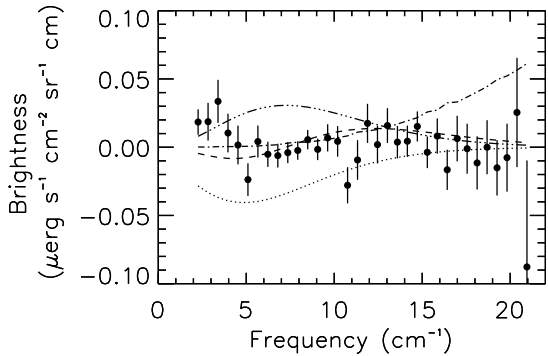

Figure 1. The data points are the *COBE* FIRAS measured CMB residuals (see text for definition). The maximum allowed 95% CL spectral distortion curves are shown for $y = 2.5 \times 10^{-5}$ (— —) and $|\mu| < 3.3 \times 10^{-4}$ ($\cdots$). Also shown is the Galaxy spectrum scaled to one fourth the flux at the Galactic pole ($\cdot$ -), and the effects of a 200 $\mu$K shift in the CMB temperature ($\cdots$ —).

The absolute temperature of the CMB is primarily needed for comparisons with different experiments, such as ground-based measurements and interstellar CN measurements. The FIRAS result is $T = 2.726 \pm 0.010$ K (95% confidence), where the uncertainty is entirely from systematic errors. The CMB temperature determines the number density of photons in the universe, $N = 8\pi\zeta(3)\Gamma(3)(kT_{CMB}/hc)^3 = 410 \pm 5$ cm$^{-3}$. The ratio of baryons to photons is $\eta = (3.95 \pm 0.25) \times 10^{-10}$ [17] so the baryon density is $(1.62 \pm 0.10) \times 10^{-7}$ cm$^{-3}$, or $\Omega_B h^2 = 0.0144 \pm 0.008$ where $h = H_0/100$ km s$^{-1}$ Mpc$^{-1}$.

## 2.3. Interpretation of CMB Spectral Observations

The interpretation of the FIRAS spectrum is given by Wright et al. [18] and summarized here. Large CMB spectrum distortions are very difficult to produce in plausible versions of the hot big bang universe. After the annihilation of positrons, the CMB energy density far exceeded the rest mass energy density of the baryonic matter until quite recently. Consequently, there are few processes involving the baryonic matter that can liberate much energy and change the CMB spectrum significantly. It is difficult to produce enough energy to create the CMB radiation from anything except a hot big bang, so the most immediate conclusion of the FIRAS measurement is that the hot big bang is the only natural explanation for a nearly perfect blackbody. (Alternatively, if the dust in intergalactic space thermalizes the radiation, then that dust must have substantial optical depth over an interval of cosmic history. That moment cannot be recent, or we would not be able to see distant galaxies at far infrared wavelengths. The *IRAS* galaxy at $z = 2.286$ demonstrates that one can see very far, and if the millimeter wave optical depth were large we would not see such an object.)

Little of the energy in the CMB was added to it after the first year of the expansion. The fraction of the CMB energy added is approximately $0.71\mu$ in the redshift range $3 \times 10^6 > z > 10^5$. For later redshifts, the fraction is $4y$. There are many possible sources of such energy augmentations, including decay of primeval turbulence, elementary particles, cosmic strings, or black holes. The growth of black holes, quasars, galaxies, clusters, and superclusters might also convert energy from other forms.

Wright et al. [18] also give limits on hydrogen burning following the decoupling. Population III stars liberate energy that is converted by dust into far infrared light (using an optical depth of 0.02 per Hubble radius). Assuming $\Omega_b h^2 = 0.015$, less than 0.6% of the hydrogen could have been burned after $z = 80$. Also, less than 0.8% of the hydrogen could have been burned in evolving IR galaxies, such as those observed by *IRAS*. *COBE* limits were obtained on the heating and reioniza-tion of the intergalactic medium. It does not take much energy to reionize the medium, relative to the CMB energy, because there are so few baryons relative to CMB photons. Even the strict FIRAS limits permit a single reionization event to occur as recently as $z = 5$. More detailed calculations [19] show that the energy required to keep the intergalactic medium ionized over long periods of time is much more substantial and quite strict limits can be obtained. If the FIRAS limits were about a factor of five more strict, then it would be possible to test the ionization state of the IGM all the way back to the decoupling. If the IGM were hot and dense enough to emit the diffuse X-ray background light, it should distort the spectrum of the CMB by inverse Compton scattering. This is a special case of the Comptonization process, with small optical depth and possibly relativistic particles. Calculations show that a smooth hot IGM could have produced less than $10^{-4}$ of the X-ray background, and that the electrons that do produce the X-ray background must have a filling factor of less than $10^{-4}$.

## 3. CMB ANISOTROPY

Years of unsuccessful searches for fluctuations in the CMB temperature placed increasingly severe upper limits on the fluctuation amplitude (see, e.g., [20,21] reviews). The lack of detectable large angular scale fluctuations was difficult to explain since regions of the sky separated by more than a couple of degrees were never in causal contact in the history of the universe, and thus had no way to establish a uniform temperature with such high precision: the *horizon problem* [22,23]. The simple big bang model takes the isotropy of the CMB as an initial condition. Also, measurements have long indicated that our local universe is nearly flat. To account for this in the big bang model the flatness must be set as an initial condition with extreme accuracy: the so-called *flatness problem* [24]. The inflation scenario [25–31] describes an early epoch when the universe expanded exponentially. In this way the horizon problem is alleviated, since our entire observable universe inflated from a small region that was in causal contact at an early epoch, and the flatness problem is solved since inflation drives the spatial curvature radius towards infinity. In principle a full theory of the early universe could predict the amplitude and spectrum of CMB temperature fluctuations, but there currently is no such generally accepted theory.

The Doppler effect due to our motion, to first order, creates an apparent dipole temperature distribution across the sky. The first firm discoveries of the CMB dipole were made by Conklin [32] and Henry [33] at the fractional temperature level of $\Delta T/T_{CMB} \sim 10^{-3}$. Later measurements [34,35] confirmed the earlier dipole discovery. Other than this simple dipole pattern, no other temperature fluctuations had been seen – the CMB radiation field appeared, in the 1980s, to be very isotropic. The high degree of isotropy supported the cosmic origin of the radiation, but spatial fluctuations were expected:

$\theta \gg 1°$ : Small gravitational potential fluctuations in the universe at the epoch when the radiation was last scattered from electrons results in temperature fluctuations. The CMB is gravitationally redshifted from primordial gravitational fluctuations, causing CMB temperature fluctuations. This "Sachs-Wolfe effect" [37] dominates on the largest angular scales, where the regions sampled are so distant from one another that there was insufficient time in the history of the universe for causal contact to occur. Two points separated by $ct$ at the decoupling, where $t$=300,000 years is the age of the universe then, now appear a few degrees apart. A domain this size at decoupling would grow by a factor of 1000 and would now be 100 Mpc across, about the size of the largest observed clustering structures.

$\theta \approx 1°$ : Electrons move during decoupling, producing Doppler shifts in the CMB temperature. The Doppler effect is expected to produce a peak temperature fluctuation at about the 0.1-1° angular scale size (i.e. spherical harmonic order $\ell \sim 200/\Omega^{0.5}$). The angle at which the Doppler peak is expected is relatively insensitive to the primordial spectral index, a cosmological constant, or the effects of gravity waves [38]. It depends primarily on the geometry of the universe, i.e the angular size of the horizon at the surface of last scattering, and thus on $\Omega_0$.

$\theta \ll 1°$ : On the smallest angular scales temperature fluctuations result from tightly coupled radiation-matter adiabatic fluctuations. Fractional changes in the mass density are proportional to fractional changes in the temperature. This effect is expected to dominate temperature fluctuations on the smallest angular scales, but the nature of these fluctuations depends on the specific cosmological content and geometry of the universe. The angle corresponding to mass $M$ at the surface of last scattering is $\Delta\theta \sim 10(\Omega h)^{2/3}(M/10^{15}hM_\odot)^{1/3}$. The imperfect coupling between matter and radiation creates a viscosity from photon diffusion, causing a decrease in perturbation amplitudes at small scales [39]. The angle subtended by the "Silk damping" scale at the surface of last scattering is described by the spherical harmonic index $\ell \sim 9 \times 10^3 \Omega^{-0.75}(\Omega_b h)^{0.5}$.

In standard big bang cosmology the fully ionized plasma in the universe recombines at $z \sim 1300$ and becomes neutral. If the universe is reionized (for example by the energy released through a burst of star formation at high redshift), then the reionization will suppress small scale CMB fluctuations. In particular, if the universe reionized at the redshift $z_R$ then CMB fluctuations are suppressed on angular scales below the horizon size at $z_R$, i.e. $\theta \sim 2(\Omega/z_R)^{0.5}$. Reionization is expected at $z \sim 50$ in $\Omega = 1$ CDM models [40]. Observations of $z \sim 4$ quasars indicate that the universe is mostly ionized to $z \sim 4$ [41]. The detection of intergalactic singly ionized helium (304 Å) "Gunn-Peterson" absorption against a $z = 3.286$ quasar also indicates an ionized intergalactic medium at $z < 3$ [42]. CMB anisotropy experiments may be able to distinguish $\Omega_0 = 1$ from $\Omega_0 < 1$ since the suppression of small scale CMB fluctuations depends almost entirely on the geometry of the universe [43].

### 3.1. COBE DMR Results

The first year of data from the *COBE* DMR instrument was used to discover and map primordial temperature fluctuations of the CMB at an angular resolution of 7° [44-46]. These results were supported by a detailed examination of the DMR calibration and its uncertainties [47] and a detailed treatment of the upper limits on residual systematic errors [48]. Bennett et al. [45] showed that spatially correlated Galactic free-free and dust emission could not mimic the frequency spectrum nor the spatial distribution of the observed fluctuations. Bennett et al. [49] also showed that the pattern of fluctuations does not spatially correlate with known extragalactic source distributions. Confirmation of the *COBE* results was attained by the positive cross-correlation between the *COBE* data and data from balloon-borne observations at a shorter wavelength [50].

Bennett et al. [36] report the results from analysis of two years of DMR flight data. The results from the two year data are consistent with those from the first year alone. The best fit dipole from the two year DMR data is $3.363 \pm 0.024$ mK towards Galactic coordinates $(\ell, b) = (264.4° \pm 0.2°, +48.1° \pm 0.4°)$ for $|b| > 15°$ [36], in excellent agreement with the first year results of the DMR [51] and with FIRAS [16]. The dipole is removed for all further analysis of the two year data, below.

The pattern of CMB fluctuations was predicted [52-54] to be scale invariant, with equal RMS gravitational potential fluctuations on all scales. A scale invariant spectrum, $P(k) \propto k^n$, where $P(k)$ is the CMB fluctuation power at comoving wavenumber $k$, is also a natural consequence of the inflationary model when $n \approx 1$. The power spectrum of the data [36,55,56] is consistent with the scale-invariant Peebles-Harrison-Zeldovich power law spectrum of primordial density fluctuations.

In general, a given cosmological model does not predict the exact CMB temperature that would be observed in our sky, but rather it will predict a statistical distribution of anisotropy parameters, such as spherical harmonic amplitudes. In the context of such models, the true CMB temperature observed in our sky is only a single realization from a statistical distribution. Thus, in addition to experimental uncertainties, we must also assign a *cosmic variance* uncertainty to cosmological parameters derived from the DMR maps. Cosmic variance can be approximately expressed as $\sigma(T_\ell^2)/T_\ell^2 \approx \sqrt{4\pi/\Omega(\ell+0.5)}$ where $\sigma$ is the rms fluctuation at an angle corresponding to the spherical harmonic order $\ell$. It is important to recognize that cosmic variance exists independent of the quality of the experiment.

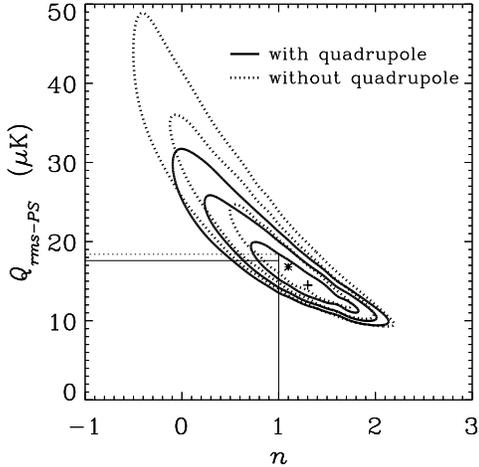

Figure 2. The likelihood function of the parameters $Q_{rms-PS}$ and $n$ based on the Bennett et al. analysis of the first two years of DMR data approximately corrected for bias effects that arise for a low-quadrupole universe.

The angular correlation function of the anisotropy is analyzed in Bennett et al. [36]. The likelihood function of the values of the cosmological parameters $n$ and $Q_{rms-PS}$ is shown in Figure 2. Bennett et al. [36] deduced $n = 1.42^{+0.49}_{-0.55}$ and $n = 1.11^{+0.60}_{-0.55}$ with and without the quadrupole, respectively, for the combined 53 GHz and 90 GHz DMR 2-year data, where the quoted uncertainties for $n$ and $Q_{rms-PS}$ encompass the 68% confidence region in two dimensions and include cosmic variance. Marginal values are $n_{marg} = 1.42\pm0.37$ and $1.11\pm0.40$, where $L(n) = max\{L(Q|n)\}$).

There is likely to be a non-zero quadrupole at a level of $Q_{rms} = 6 \pm 3$ $\mu$K (68% CL). $Q_{rms}$ has a lower value than the quadrupole expected from a fit to the entire power spectrum, $Q_{rms-PS}$, but whether this is due to cosmic variance, Galactic model error, or reflects the cosmology of the universe remains to be determined. The probability of measuring a quadrupole of amplitude $3 < Q_{rms}$ ($\mu$K) $< 9$ from a power spectrum normalized to $Q_{rms-PS} = 17$ $\mu$K is 10%.

Bennett et al. selected from a sample of simulated maps the subset of maps in which the actual quadrupole moment was close to the low value observed in our sky (between 3 and 9 $\mu$K). The resulting subset of most-likely $n$ values had a median of $+1.31 \pm 0.04$ so there is a bias of $+0.31 \pm 0.04$ in $n$ for the low quadrupole case. Correcting for a bias of 0.31 they deduced $n = 1.22^{+0.49}_{-0.55}$ (or marginal value $n_{marg} = 1.22\pm0.37$), in good agreement with the results obtained excluding the quadrupole.

It is also possible to analyze the maps in terms of spherical harmonics. There is considerable subtlety and difficulty in this since the removal of the galactic plane causes the harmonics to be non-orthogonal, and produces strong correlations among the fitted amplitudes. Figure 3, adapted from Wright et al. [55], is a power spectrum analysis using modified spherical harmonics. The figure also shows the results of other instruments. A scale-invariant power spectrum would produce a horizontal line on this plot. Górski et al. [56] derive the power spectrum using orthogonal polynomials on the galactic cut sphere. Power spectral results from the first two years of *COBE* DMR data are summarized in Table 1. (Note that Górski et al. define $n_{marg}$ differently than [36] and [55], using $L = \int L(Q_{rms-PS}, n) dQ_{rms-PS}$).

The FIRAS data provide a limit on the spectral index of primordial density fluctuations. Wright et al. [18] found an upper limit of $n < 1.9$, based on the work of Daly [57,58]. Hu, Scott, & Silk [59] claim an upper limit of $n < 1.7$, assuming that the primordial mass scale extends to 1 $M_\odot$. These calculations do not significantly limit $n$ beyond the direct measurements.

It is clear from the variety of models that there can be no unique interpretation of the spectral index or the amplitude of the fluctuations. Additional data will be needed to establish a single picture of the early universe.

It is important to determine whether the pri-mordial fluctuations are Gaussian. The raw distribution of the temperature residuals should be close to Gaussian if the sky variance is Gaussian and the receiver noise is Gaussian. The receiver noise varies somewhat from pixel to pixel because the observation times are not all the same, but when this is taken into account the data appear Gaussian [60]. There is no evidence that there is an excess of large deviations, as would be expected if there were an unknown population of point sources. A search for point sources [61] in the two-year maps was negative. Given the large beam of the instrument and the variance of both cosmic signals and receiver noise, it is still possible for interesting signals to be hidden in the data. The three point correlation function provides an excellent test of the statistical properties of the maps. Hinshaw et al. [62] computed the equilateral three-point correlation function of the DMR maps and the results are consistent with Gaussian statistics. Tests of the DMR maps using the genus statistic produce results that are also consistent with Gaussian statistics [60]. While no evidence for deviations from Gaussian statistics were found, most popular alternative theories of cosmic structure produce near-Gaussian statistics on large angular scales.

Table 1
Power Spectrum Analysis Results of Combined 53 GHz and 90 GHz DMR 2-yr Data

| parameter | incl $\ell=2$ | 2-pt corr func Bennett et al. [36] | modified $Y^\ell_m$ Wright et al. [55] | orthog poly. Górski et al. [56] |
|---|---|---|---|---|
| $n$ | Y | $1.42^{+0.49}_{-0.55}$ | $---$ | $1.22^{+0.43}_{-0.52}$ |
| $n_{marg}$ | Y | $1.42 \pm 0.37$ | $1.39^{+0.34}_{-0.39}$ | $1.10 \pm 0.32$ |
| $Q_{rms-PS}$ ($\mu$K) | Y | $14.3^{+5.2}_{-3.3}$ | $---$ | $17.0^{+7.6}_{-4.8}$ |
| $Q_{rms-PS\|n=1}$ ($\mu$K) | Y | $18.2 \pm 1.5$ | $---$ | $19.9 \pm 1.6$ |
| $n$ | N | $1.11^{+0.60}_{-0.55}$ | $---$ | $1.02^{+0.53}_{-0.59}$ |
| $n_{marg}$ | N | $1.11 \pm 0.40$ | $1.25^{+0.40}_{-0.45}$ | $0.87 \pm 0.36$ |
| $Q_{rms-PS}$ ($\mu$K) | N | $17.4^{+7.5}_{-5.2}$ | $---$ | $20.0^{+10.5}_{-6.5}$ |
| $Q_{rms-PS\|n=1}$ ($\mu$K) | N | $18.6 \pm 1.6$ | $19.8 \pm 2.0$ | $20.4 \pm 1.7$ |
| $a_\ell(\mu K)$ indep of $n$ | $-$ | $a_7 \cong 9.5$ | $---$ | $a_9 \cong 8.2$ |

### 3.2. Other CMB Anisotropy Observations

A large number of experiments are being done from the ground and using balloons with a beam size of $\sim 0.5°$. Most of these groups are reporting detections of anisotropy at the $\sim 10^{-5}$ level. This angular scale does not probe unmodified primeval fluctuations, so calculation and interpretation are necessary to connect the data with the COBE data and galaxy clustering data.

The Far Infra-Red Survey (**FIRS**) [50,63–65] is a balloon-borne anisotropy experiment with observations covering most of the Northern sky. The maximum likelihood cosmological parameters from the FIRS data are $n = 1.0$ and $Q = 19$ $\mu$K. A cross-correlation between the FIRS data and the COBE DMR data shows that the anisotropy seen in each experiment has a common spatial origin. Given the differing observing frequencies of the two experiments, this positive cross-correlation lends strong support to the cosmic interpretation of the observed fluctuations.

The Medium Scale Anisotropy Measurement (**MSAM**) balloon-borne experiment [66] made the first detection of medium-scale CMB anisotropy. The Millimeter-wave Anisotropy eXperiment (**MAX**) is also a balloon-borne package [67–71] that has completed four flights that saw significant fluctuations in the CMB. All of the MAX fields have consistent levels of anisotropy except for the Mu Pegasus scan observed on the third flight of MAX. The Advanced Cosmic Microwave Explorer **ACME South Pole** [72,73] reported upper limits on fluctuations using the same optical configuration as MAX, but operating with high electron mobility transistor (HEMT) amplifiers. **ULISSE** [74] reported up-per limits on 6° CMB anisotropy using millimeter balloon borne bolometric observations. The **ARGO** [75] balloon- borne experiment observed a statistically signal with a 52' beam.

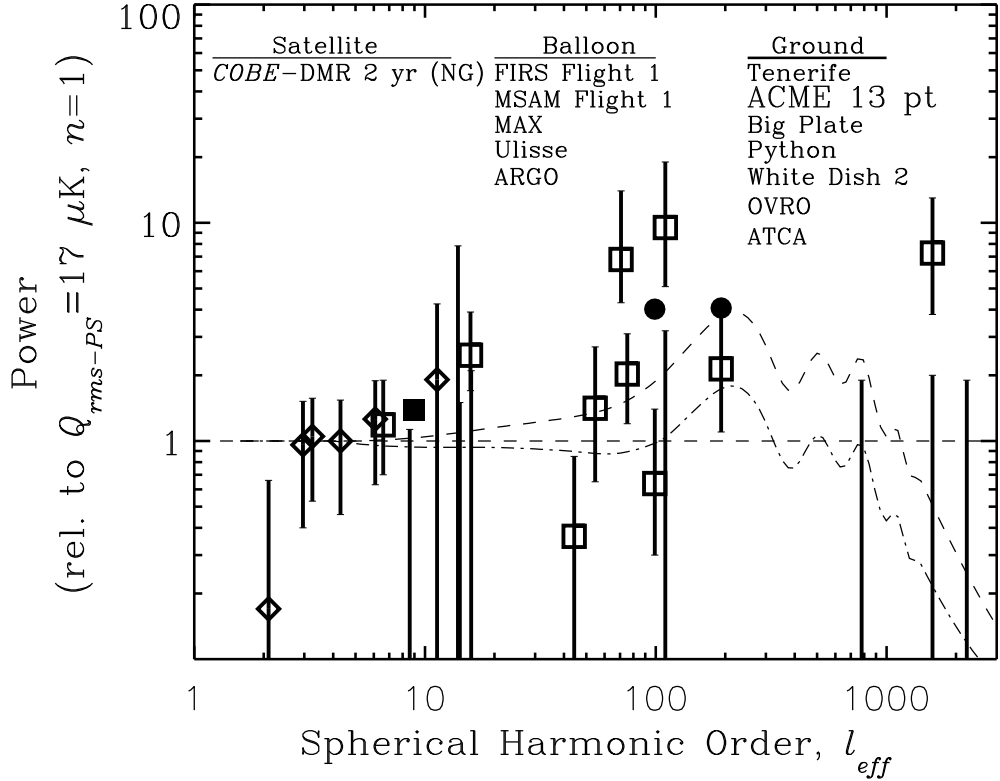

Figure 3. Power spectra for spherical harmonic number $\ell_{eff}$. The diamond points on the left are Galaxy-removed COBE DMR results from Wright et al., and the small box at $\ell = 9$ is the normalization derived from the use of the overall COBE power spectrum by Górski et al. The two filled circles show the effects of not removing sources from the MSAM data. The two dashed curves are indicative of power spectra expected in dark matter models.

The "**Big Plate**" experiment [76] used HEMT amplifiers to detect of CMB anisotropy from Saskatoon, SK, Canada. CMB fluctuations were also detected and mapped from ground-based radiometers in **Tenerife** [77]. Fluctuations were reported from South Pole observations by **Python** [78]. Also from the South Pole **White Dish** [79] reports an upper limit on CMB anisotropy. Arc-minute scale anisotropy limits were reported using the Owens Valley Radio Observatory (**OVRO**) [80]. The potential for radio source contamination at 20 GHz on small angular scales is a major problem. The Australia Telescope Compact Array **ATCA** was used to place upper limits on CMB anisotropy in a Fourier synthesized image [81].

## 4. COMMENTS ON CMB MEASUREMENTS

COBE FIRAS made a precision measurement of the spectrum of the CMB, providing strong support for a simple big bang model. Large angular scale CMB measurements are consistent with the expected $n = 1$ power-law spectrum. (If the effects of a standard cold dark matter model are included, *COBE* DMR should find $n \approx 1.05$ for a $n = 1$ universe.) On smaller angular scales all cosmological models dominated by nonbaryonic dark matter predict a Doppler peak. Calculations of CMB anisotropies in open CDM-dominated universes with adiabatic primordial density perturbations and differing reionization histories show that CMB anisotropies depend almost exclusively on the geometry of the universe [43]. The location of the Doppler peak depends on $\Omega$ ($\ell \sim 200/\Omega^{1/2}$) and is fairly insensitive to the specific values of the baryonic mass density, the Hubble constant, or the cosmological constant. Reionization decreases the amplitude of the Doppler peak by approximately $e^{-2\tau}$ where the optical depth $\tau \approx 0.04\Omega_b h \Omega^{-1/2} x_e (z_{ls})^{3/2}$ and $\Omega_b$ is the mass density in baryons, $z_{ls}$ is the redshift of the last scattering surface, and $x_e$ is the ionization fraction. $\tau$ is the reionization optical depth so $\tau = 0$ means no reionization. If the Doppler peak amplitude at $\ell \sim 200$ is greater than at $\ell \sim 800$ then an $\Omega = 1$ universe is suggested. Alternately, if the amplitude at $\ell \sim 800$ is greater than at $\ell \sim 200$, an $\Omega < 1$ universe is suggested. Has a CMB anisotropy Doppler peak been detected in our Universe? While some claim it has, we remain unconvinced by the existing set of observations. Fortunately, the future looks bright for obtaining more definitive data. Several groups are planning long duration balloon flights and other groups are planning for future space missions. It is because the CMB is an important and unique probe of our universe that such extraordinary efforts are justified.